\documentclass[aps,prd,twocolumn,showpacs,amsmath,amssymb,nofootinbib]{revtex4}

\usepackage{epsfig}
\usepackage{graphicx}
\usepackage{amsfonts}
\usepackage{amssymb}
\usepackage{bm}
\usepackage{latexsym}
\usepackage{amsmath}
\usepackage{hyperref}
\usepackage{xcolor}
\usepackage{multirow}
\usepackage{dcolumn}

\newcommand{\beq}{\begin{equation}}
\newcommand{\eeq}{\end{equation}}
\newcommand{\bea}{\begin{eqnarray}}
\newcommand{\eea}{\end{eqnarray}}

\begin{document}


\title{A Unified Universe}

\author{Alessandro Codello\footnote{E-mail:~codello@cp3-origins.net}}
\author{Rajeev Kumar Jain\footnote{Current address: Department of Physics, Indian Institute of Science, Bangalore 560012, India. E-mail: rkjain@iisc.ac.in}}
\affiliation{CP$^3$-Origins, Centre for Cosmology and Particle Physics Phenomenology, \\
University of Southern Denmark, Campusvej 55, 
5230 Odense M, Denmark}

\begin{abstract}

We present a unified evolution of the universe from very early times until the present epoch by including both the leading local correction $R^2$ and the leading non--local term $R\frac{1}{\square^2}R$ to the classical gravitational action.
We find that the inflationary phase driven by $R^2$ term gracefully exits in a transitory regime characterized by coherent oscillations of the Hubble parameter. The universe then naturally enters into a radiation dominated epoch followed by a matter dominated era. At sufficiently late times after radiation--matter equality, the non--local term starts to dominate inducing an accelerated expansion of the universe at the present epoch. We further exhibit the fact that both the leading local and non--local terms can be obtained within the covariant effective field theory of gravity.  This scenario thus provides a unified picture of inflation and dark energy in a single framework by means of a purely gravitational action without the usual need of a scalar field.
\end{abstract}
\date{\today}
\maketitle

\paragraph*{\bf Introduction.}
Cosmological observations strongly support the idea that the universe underwent an early period of accelerated expansion called inflation \cite{Guth:1980zm,Ade:2015lrj}. Besides, local supernovae measurements \cite{Riess:1998cb,Perlmutter:1998np} also suggest that the universe is experiencing a phase of acceleration at the present epoch, caused by dark energy \cite{Peebles:2002gy}. Whether there exists a deeper and fundamental connection between the two (or not), an interesting question to ask is if it is possible to unify both these epochs in a single framework with minimal degrees of freedom. Such scenarios have been explored by using matter fields (scalar fields with an appropriate potential) as well as by modifying gravity.
Within General Relativity (GR) and in the absence of a cosmological constant, it is not possible to explain either inflation or dark energy without adding extra degrees of freedom and therefore, these two epochs are very novel consequences of physics beyond classical GR described by the Einstein--Hilbert (EH) action.
The framework of the covariant Effective Field Theory (EFT) of quantum gravity, developed in \cite{Codello:2015mba, Codello:2015pga,Codello:2016xhm}, predicts both local and non--local correction terms which then become natural candidates for driving inflation and dark energy thereby allowing us to construct a unified picture of the universe. The key advantage of using EFT methods is that they allow to compute such corrections from first principles even in the absence of a complete theory of quantum gravity \cite{Donoghue:1994dn,Burgess:2003jk,Donoghue:2012zc,Donoghue:2015hwa}.
\begin{figure*}[t]
	\begin{center}
		\includegraphics[width=1.01\linewidth]{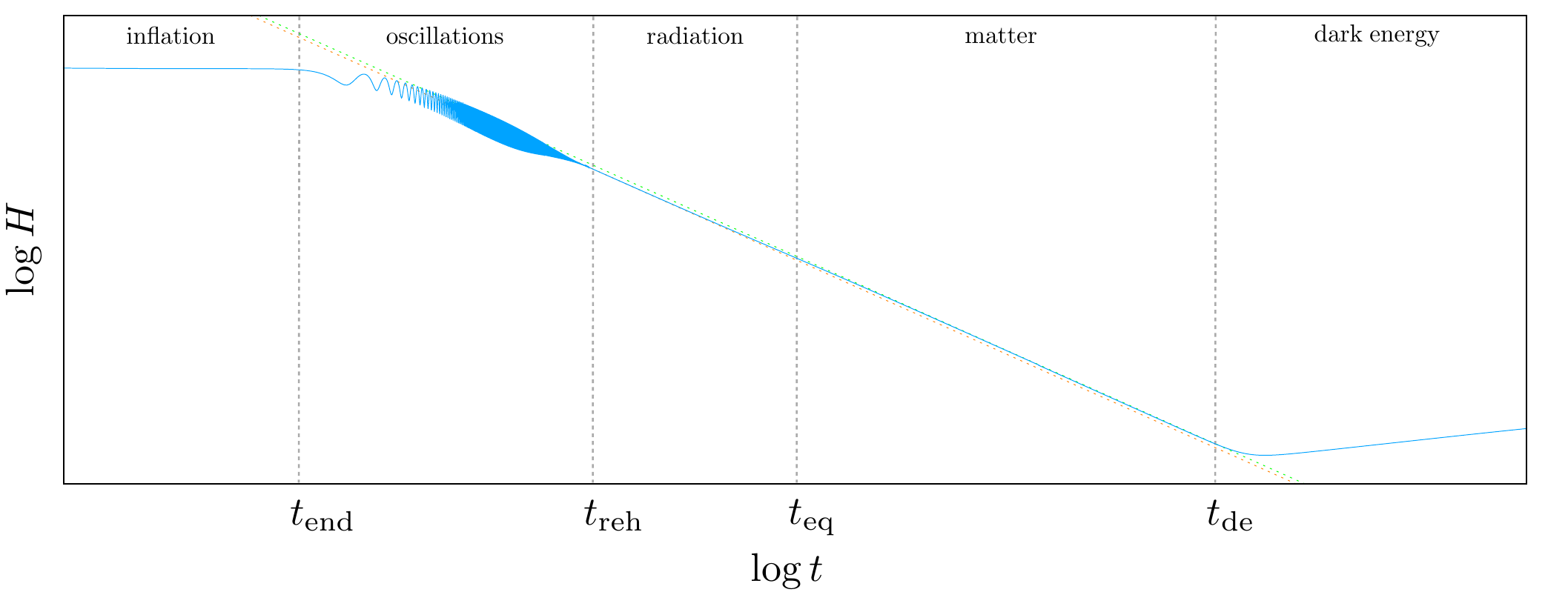}
		\caption{
		A unified evolution of the universe from very early times until today.
		We plot $\log H$ vs. $\log t$ starting from an inflationary stage, passing through a transitionary phase characterized by strong coherent oscillations, followed by radiation and matter to finally end in a dark energy dominated regime. The critical epochs that separate these phases are end of inflation $t_{\rm end }$, reheating $t_{\rm reh}$, radiation--matter equality $t_{\rm eq}$ and the onset of dark energy $t_{\rm de}$.
\label{universe}}
	\end{center}
\end{figure*}

In addition to scalar fields, an inflationary epoch in the early universe can also be driven by higher order curvature corrections to the EH action, notably by a $R^2$ term -- a scenario first proposed by Starobinsky \cite{Starobinsky:1980te} and further discussed in \cite{Mukhanov:1981xt,Starobinsky:1983zz,Kehagias:2013mya}. This scenario is one of the simplest and oldest model of inflation based on purely gravitational corrections. Albeit the existence of a large number of inflationary models,  Starobinsky inflation should be one of the most probable among all from an Occam's razor point of view and also turns out to be the most preferred model with the highest Bayesian evidence in the recent datasets \cite{Martin:2013tda,Ade:2015lrj}. The presence of higher order curvature  terms (including the derivatives) can be understood by means of one--loop quantum corrections to the EH action as suggested in \cite{Starobinsky:1980te} and extended, for example, in \cite{Codello:2014sua}. In the covariant EFT of gravity, we found that a $R^2$ term naturally arises as a leading local correction to the EH action and becomes responsible for driving inflation at early times without the need of additional matter fields. Thus, inflation in this set--up is entirely a feature of the leading corrections in the gravitational sector \cite{Codello:2015pga}.

Within the framework of GR, cosmic acceleration at the present epoch can be achieved by adding a cosmological constant $\Lambda$ and is indeed the simplest possibility. Recently, it has been realized that specific curvature square non--local terms can also drive the current acceleration of the universe. In particular, the non--local term $R \frac{1}{\square^2} R$ leads to a very interesting phenomenology as it effectively behaves like a cosmological constant as $ R/\square \to 1$ at late times. This term has been argued as a consistent IR modification of GR and the theory together with the EH action remains free of any propagating ghost--like degree of freedom or other instabilities \cite{Maggiore:2014sia}. This scenario provides a viable alternative to dynamical dark energy and  has been greatly studied including the study of cosmological perturbations and other observable imprints \cite{Dirian:2014ara, Maggiore:2015rma}.
In order to understand the origin of such non--local terms in a consistent framework, we have recently shown that various non--local terms appear at the second order in a curvature expansion \cite{Codello:2015mba, Codello:2015pga}. However, the $R \frac{1}{\square^2} R$ term among others is the most relevant one for dark energy due to being roughly a constant at late times. Lately, modifications of GR including the Weyl--square term, also predicted by the covariant EFT \cite{Codello:2015mba}, have also been studied. Such terms do not contribute to the  background expansion but only to the evolution of cosmological perturbations \cite{Cusin:2015rex}.

In this Letter, we present a unified evolution of the universe from very early times until the present epoch by including both the leading local correction $R^2$ and the leading non--local contribution $R\frac{1}{\square^2}R$ to the EH action together with radiation and matter.
We find that the initial inflationary epoch induced by the $R^2$ term exists into a transitory regime with coherent and damped oscillations of the Hubble parameter and could be characterized as reheating in this scenario. This phase then naturally enters into a radiation dominated epoch followed by a matter dominated era. At sufficiently late times, the non--local term becomes dominant and leads to a dark energy dominated universe with equation of state $w\leq -1$. This set--up provides a unified picture of our universe from earliest epochs until today thereby unifying both inflation and dark energy in a single framework by means of purely gravitational corrections to the EH action. Later we will show that these corrections are naturally present in the covariant EFT of gravity.
\vskip 4pt
\paragraph*{\bf A unified scenario.}\label{unified}
%

We are interested in studying the evolution of the universe in a scenario wherein the leading local quadratic term is combined with the most relevant IR non--local terms. The total effective action including the matter part $S_{\textrm{m}}$ is then given by
\begin{eqnarray}
\Gamma=
\int {\rm d}^{4}x\sqrt{-g} \left[\frac{M^2_{\rm Pl}}{2} R
- \frac{1}{\xi} R^{2}
+ m^{4}R \frac{1}{\square^2} R\right] + S_{\textrm{m}}\,,\qquad
\label{action}
\end{eqnarray}
where $\xi$ and $m$ are phenomenological parameters which must be fixed by observations. The normalization of CMB spectrum fixes $\xi \sim 1.2\times10^{-9}$ and the dark energy density today leads to $m \sim 0.3 \sqrt{H_{0} M_{\rm Pl}}$. 
The specific non--local term $R \frac{1}{\square^2} R$ has been chosen since it is the simplest that can effectively emulate a cosmological constant term since $R \frac{1}{\square^2} R \sim 1$  as $ R/\square \to 1$ at late times.
This term as an alternative to dark energy was first proposed in  \cite{Maggiore:2014sia}.
The effective action in (\ref{action}) is the minimal unified action (with deviations from GR of purely gravitational character) capable of describing inflation and dark energy together in a single framework {\it sans} a cosmological constant was proposed in \cite{Maggiore:2015rma,Codello:2015pga,Cusin:2016nzi}.

In order to understand the implications of this effective action for our universe, we work in a $(3+1)$--dimensional, spatially flat, FRW spacetime described by the line element $ds^{2}=-dt^{2}+a^{2}(t)d{\bf x}^{2}$ where $a(t)$ is the scale factor.
Einstein's equations can now be written as $G_{\mu\nu}+\Delta G_{\mu\nu} = T_{\mu\nu}/M^2_{\rm Pl}$, where $\Delta G_{\mu\nu}$ corresponds to the correction terms arising from the local and non--local terms in (\ref{action}). It is evident that $\Delta G_{\mu\nu}$ is covariantly conserved and its explicit form can be found in  \cite{Codello:2015pga}.
The modified Friedmann equations of motion (EOM) can be obtained by varying this action
%
\begin{eqnarray}
H^{2}-\frac{12 }{M^2}\left(2H\ddot{H}+6H^{2}\dot{H}-\dot{H}^{2}\right)\hspace*{25pt} \nonumber\\
-\frac{4\, m^4}{M^2_{\rm Pl}}\left(2 H^{2}S+H {\dot S}+\frac{1}{2} {\dot H} S-\frac{1}{6} {\dot U}{\dot S}\right)
&=&
\frac{\rho}{3M^2_{\rm Pl}} \label{ST_2.1}\nonumber \\
\ddot U+3H\dot U -6\left(2H^2+\dot H\right)& = & 0  \label{ST_2.2}\nonumber\\
\ddot S+3H\dot S -U& = & 0\,, \label{ST_2.3}
\end{eqnarray}
%
where $H={\dot a}/a$ is the Hubble parameter and the two auxiliary fields $U$ and $S$ are defined as, $U=\frac{1}{-\square}R$ and $S=\frac{1}{\square^{2}}R$, respectively.
We also define the mass scale characterizing inflation as $M^2\equiv \xi M^2_{\rm Pl}$. 
%
\begin{figure}[t]
\begin{center}
\includegraphics[width=1.04\linewidth]{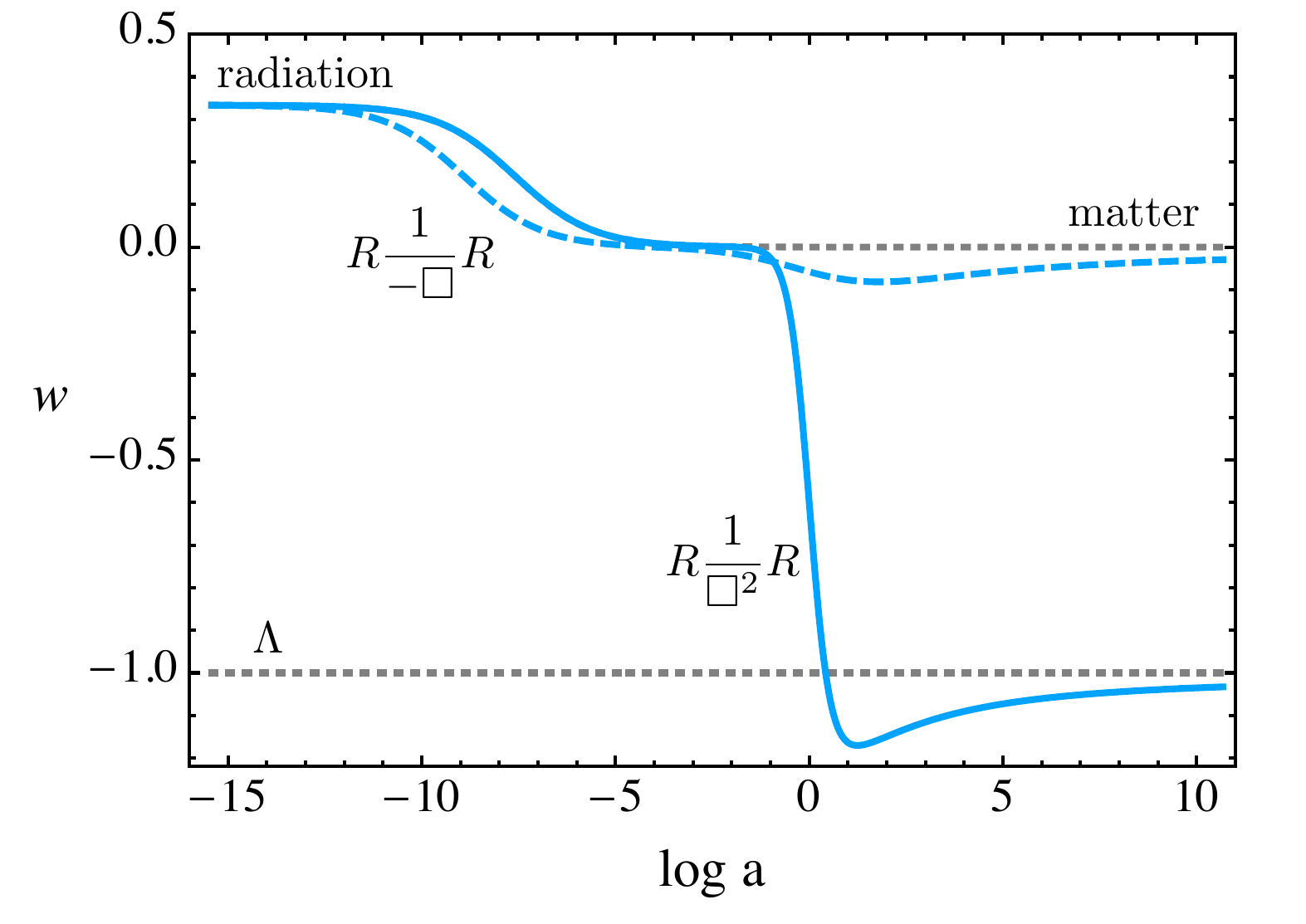}
\caption{We plot the total equation of state parameter $w$ vs. $\log a$ from radiation epoch until today. The imprints due to the first two non-local terms $R\frac{1}{-\square}R$ and $R\frac{1}{\square^2}R$ in the effective action are compared. It is evident that $R\frac{1}{\square^2}R$ effectively leads to $w\simeq -1$ at the present epoch. \label{w}}
\end{center}
\end{figure}
%
%
The modified Friedmann EOM must be solved together with the continuity equations for radiation and matter which are given by 
$\dot \rho_{{\rm i}}+3H(1+w_{\rm i}) \rho_{{\rm i}}=0,
$ where ${\rm i}=\{{\rm r},{\rm m}\}$ with $w_{\rm r}=1/3$ and $w_{\rm m}=0$.
%
At early times when local corrections are most relevant, our scenario reduces to the Starobinsky model and naturally describes inflation.
To achieve sufficient e--folds of inflation, the time derivate of the Hubble parameter at an initial time can be appropriately tuned. 
The CMB normalization instead determines the value of the coupling $\xi$. 
Note that, as pointed out in \cite{Codello:2015pga}, $R+R^2$ gravity is not exactly solvable due to the non--linear nature of the equations. However, at sufficiently early times when $R^2$ is dominant over the EH term, the theory admits a quasi de-Sitter solution leading to Starobinsky inflation. This solution can be understood by neglecting higher derivative terms in $H$ (i.e. $|\dot H| \ll H^2$) in the first equation in eqs. (\ref{ST_2.3}), leading to $H^2-\frac{72}{M^2} H^2 \dot H \simeq 0,$ which clearly admits an exponential solution for the scale factor, as expected.
This behavior is depicted by the constant $H$ solution on the left in Fig. \ref{universe}. 
Note that, the universe in this scenario {\it gracefully} exits from inflation and enters into an oscillatory regime. We obtain the oscillatory solution for the Hubble parameter numerically but one can also gain an analytical understanding of their origin by ignoring some higher order terms in the modified Friedmann EOM \cite{Codello:2015pga}. We stress that this transient oscillatory regime is very novel and generic feature of our scenario which can be considered as the reheating epoch in this set--up. This could in principle have interesting observable imprints. We leave the study of the details of the reheating stage and its possible imprints to future work.
\begin{figure}[t]
\begin{center}		
\includegraphics[width=1.04\linewidth]{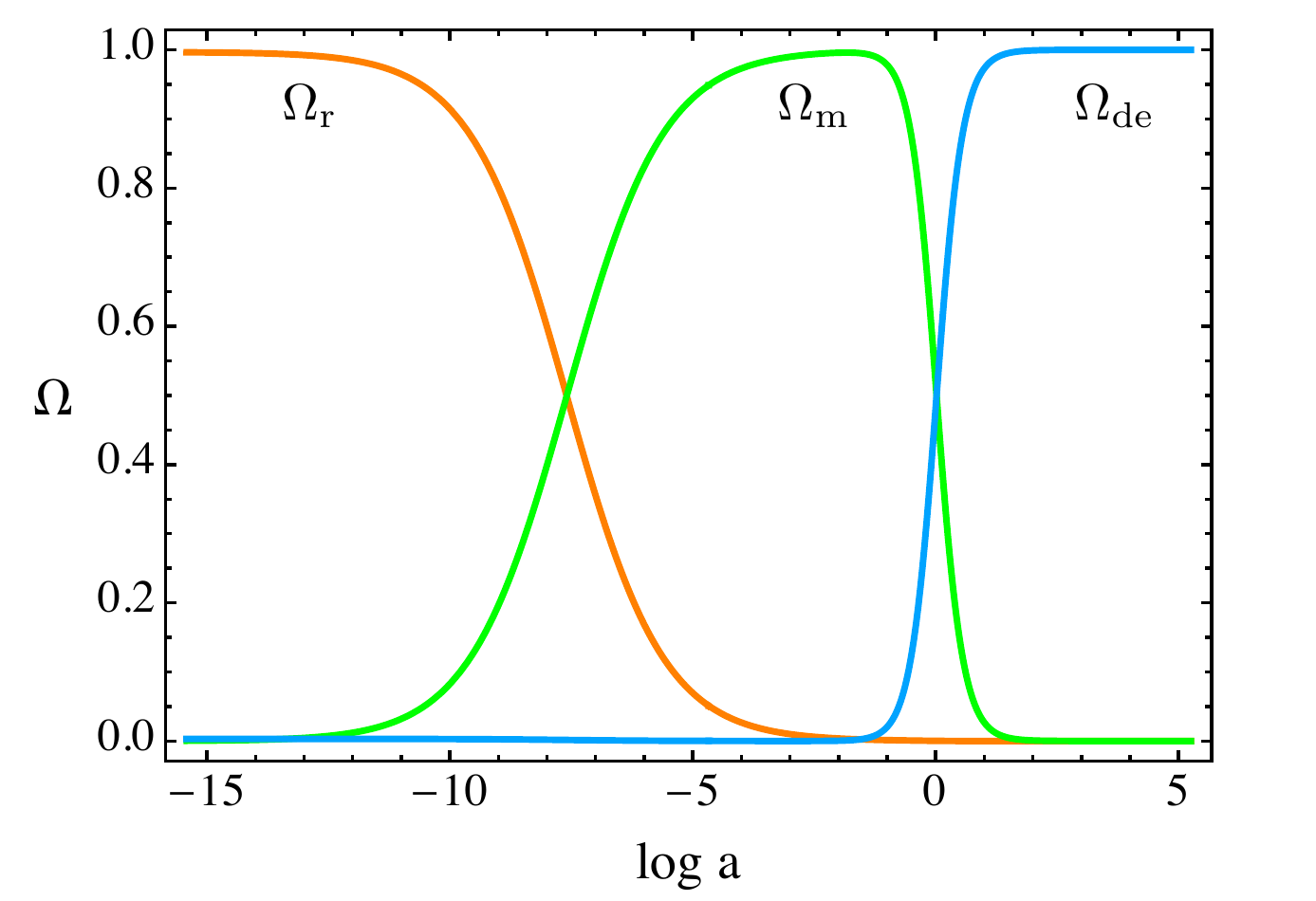}
\caption{We plot the dimensionless energy density ratio $\Omega_{\rm r},\Omega_{\rm m}$ and $\Omega_{\rm de}$ vs $\log a$. All the salient features of Fig. \ref{w} are clearly evident here. \label{omega}}
\end{center}
\end{figure}

After the oscillatory phase is over, the universe naturally enters into the radiation dominated epoch.  After radiation--matter equality, the universe then evolves in a matter dominated era as clearly shown in  Fig. \ref{universe}. At sufficiently late times after  equality, the non--local term $R\frac{1}{\square^2}R$ which was so far subdominant starts to become relevant and drives the current acceleration of the universe, due to the fact that $R\frac{1}{\square^2}R \to 1$ as $R/\square \to 1$ at the present epoch. The effects of this non--local term are studied by solving the modified Friedmann equation numerically, as shown in Fig. \ref{universe} and also discussed in \cite{Maggiore:2014sia}.

Moreover, we are also interested in understanding the imprints of the non--local term using an analytical approach in the background of both radiation and matter.
It is useful to note that in terms of the conformal time $\tau=\int dt/a(t)$, the Friedmann equation including only radiation and matter
can indeed be solved exactly and the solution is  
$a(\tau)/a_{\rm eq}=(2\sqrt{2}-2) \left(\tau/\tau_{\rm eq}\right)+(1-2\sqrt{2}+2) \left(\tau/\tau_{\rm eq}\right)^2$
with $\tau_{\rm eq} =(2\sqrt{2}-2)/a_{\rm eq} \sqrt{3 M_{\rm Pl}^2/\rho_{\rm eq}}$. 
Here, $\rho_{\rm eq}$, $a_{\rm eq}$ and $\tau_{\rm eq}$ are the energy density, scale factor and conformal time at the epoch of radiation--matter equality, respectively. 
Using this background solution in a closed form, we have solved the Friedmann EOM corresponding to the non--local term with initial conditions $U=S=0$ and $U'=S'=0$. We have also treated the contributions from the non--local term when taken on the right hand side of the EOM as an effective dark energy density.
In Fig. \ref{w}, we have plotted the total equation of state parameter $w$ as a function of $\log a$. It is evident from the figure that the universe starting from radiation epoch ($w=1/3$) smoothly transits to the matter dominated regime ($w=0$) which then evolves in the present epoch dominated by the non--local term with $w \leq -1$.
%
For later discussion we have also shown the corresponding behavior for the leading non--local term $R\frac{1}{-\square}R$ but found that only $R\frac{1}{\square^2}R$ leads to an accelerating universe today.
Moreover, in Fig. \ref{omega}, we have plotted the dimensionless energy density ratio $\Omega_{\rm r}$, $\Omega_{\rm m}$ and $\Omega_{\rm de}$ as a function of $\log a$ which perfectly corroborates the behavior of Fig. \ref{w}. By demanding $\Omega_{\rm de}\simeq0.68$ to be consistent with observations today, we can also obtain the value of the mass scale $m$ of the non--local term to be $m \sim 0.3 \sqrt{H_{0} M_{\rm Pl}}
\sim 5.7 \times 10^{-4}\, {\rm eV}$.
\vskip 4pt
\paragraph*{\bf Covariant EFT of gravity.}\label{eftg}

As we have seen, the effective action in (\ref{action}) is very appealing from a phenomenological perspective since it can naturally describe all the different phases through which the universe has evolved, but can it have a deeper origin or interpretation?
Under which assumptions is it a good low energy description of quantum gravity?
We will discuss here how the model (\ref{action}) can be justified from low energy quantum gravity when this is treated by means of EFT methods.

Since we are interested in studying the evolution of the whole universe, the EFT approach must be developed in a covariant way
in order to derive an effective action valid on an arbitrary spacetime and in particular FRW.
%
In our previous paper \cite{Codello:2015mba}, we have computed the leading order EFT action to the second order in the curvatures.
The final result for the gravitational part of the effective action is given by
%
\begin{equation} \label{ea_R2_Final}
\!\!\!\! \Gamma = \int\! {\rm d}^{4}x  \sqrt{-g}\left[\frac{M^2_{\rm Pl}}{2}R-\frac{1}{\xi} R^{2}
- R\, \mathcal{F}\!\left(\frac{-\square}{\widetilde m^2}\right)\!R \right]\,,
\end{equation}
%
where 
the structure function $\mathcal{F}$ is completely determined once the matter content of the theory is specified.
The general form of the structure function can be found in \cite{Codello:2015mba}. 
The action depends on two parameters $\xi$ and $\widetilde m$.
The first is free to tune while the second is in principle related to some mass scale of the underlying theory.
%
On FRW the Weyl tensor vanishes identically and therefore we have not reported the Weyl part of the
effective action (\ref{ea_R2_Final}) since it will not contribute to the background EOM. 
However, this is not true when dealing with cosmological perturbations and the contributions due to these terms
must be taken into account which could lead to very distinct signatures in the cosmological observables \cite{Cusin:2015rex}.

The structure function $\mathcal{F}$ is non--local in the low energy limit $\widetilde m^2 \ll -\square$ and has the following form \cite{Codello:2015mba}
%
\begin{eqnarray}
\mathcal{F}\!\left(\frac{-\square}{\widetilde m^2}\right)  &=&  \alpha\log\frac{-\square}{\widetilde m^{2}} +\beta\frac{\widetilde m^{2}}{-\square} \nonumber\\
&&\quad+\,\gamma\frac{\widetilde m^{2}}{-\square}\log\frac{-\square}{\widetilde m^{2}}+\delta\!\left(\frac{\widetilde m^{2}}{-\square}\right)^2+...
\label{F}
\end{eqnarray}
%
where the coefficients $\alpha,\beta,\gamma$ and $\delta$ are indeed {\it predictions} of the EFT of gravity which ultimately depend only on the field content of the theory\footnote{They are of order $N$ where $N$ is the number of particles of a given species.}  and are listed in \cite{Codello:2015mba}.
%
We recognize in (\ref{F}) the appearance of the non--local term $R\frac{1}{\square^2}R$ assumed in the model (\ref{action}) if $m \equiv |\delta|^{\frac{1}{4}}\widetilde m$.
As mentioned earlier, for viable cosmology, we should have $m^2 \sim {H_{0} M_{\rm Pl}}$. Since $\square^{-1} \sim H_0^{-2}$ at the present epoch, the low energy limit $\widetilde m^2 \ll -\square$ translates to $|\delta|^{\frac{1}{2}}\frac{H_{0}}{M_{\rm Pl}} \gg 1$. Although $H_{0} \ll M_{\rm Pl}$, this inequality can still be satisfied for a suitable choice of $\delta$ or $N$ as $\delta \propto N$.  These relations also imply that  $|\delta|^{\frac{1}{2}} \widetilde m^2 \sim {H_{0} M_{\rm Pl}} \sim 10^{-6}\, {\rm (eV)^2}$ and $\widetilde m \ll {H_{0}} \sim 10^{-33}\, {\rm eV}$. Therefore, the low energy expansion can be justified for a suitable value of $\delta$ for the structure function $\mathcal{F}$. 
Even though there exists specific matter choices
that can make one or more of these coefficients vanish, in the most general case these are all non--zero and so all other non--local terms are in principle present in the effective action \cite{Codello:2015pga}.

Non--local modifications of GR based on one or more of the operators in (\ref{F}) have been considered by many authors, see for example  \cite{Espriu:2005qn,Deser:2007jk,Nojiri:2007uq,Koivisto:2008xfa,Nojiri:2010pw,Biswas:2010zk,Nojiri:2010wj,Zhang:2011uv,Modesto:2013jea,Donoghue:2014yha,Modesto:2014lga,Zhang:2016ykx} and references therein.
In order to justify the model (\ref{action}) we need to show that indeed the operator $R\frac{1}{\square^2}R$ dominates the late time evolution of the universe.
For this reason in Fig. \ref{w} we have compared the equation of state parameter for a dark energy fluid generated by the first two non--local terms $R\frac{1}{-\square}R$ and $R\frac{1}{\square^2}R$. As can be seen, only the latter is capable of changing the late time evolution while the first only affects the intermediate regime.
The terms containing the $\log$ operators are more delicate to define and have been discussed in \cite{Codello:2015mba}. 
The analysis shows that their behavior is sub--leading with respect to the relative operator without the $\log$ and so to first approximation can be discarded.
The $R \log \frac{-\square}{m^2} R$ term may be relevant at early times but in our model this epoch is dominated by the local $R^2$.
Finally, due to the smallness of the mass scale $m$, all higher order operators will be suppressed by large additional factors and to first approximation can also be discarded.
Of the terms contained in the expansion (\ref{F}) the one dominant at late times is thus the one included in equation (\ref{action}) and the 
effective action (\ref{action}) is in principle justifiable from first principles via EFT arguments.
%
%
%

As discussed in \cite{Codello:2015mba}, the effective action also contains many new non--local operators with three or more curvatures, as for example $R\frac{1}{-\square}R\frac{1}{-\square}R$.  Within these there is also the contribution from the conformal anomaly.
These operators may have interesting cosmological implications (see for instance \cite{Shapiro:2008sf,Mottola:2010gp} and references therein) and, even if sub--leading when compared with the two curvature terms considered in this Letter, their effects should be studied carefully, 
for a recent discussion, see \cite{Netto:2015cba,Cusin:2016nzi}. 

Finally, we can try to answer the question: what is $m$? In the case of scalars and fermions, $m$ is the effective mass of the particle.
For these cases the mass required for observational consistency is $m \sim 0.57\, {\rm meV}$ which makes it a very light particle. Photons do not  contribute while gravitons can induce an effective gravitational mass $m^2_{\rm grav}=2\, V(v)/M^2_{\rm Pl}$ (where $V(\varphi)$ is the scalar potential with $v$ its minimum). Such a light particle must have been relativistic all its life during the evolution of the universe and would contribute to the effective number of relativistic degrees of freedom $N_{\rm eff}$. For instance, a light scalar in equilibrium with the same temperature as neutrinos would lead to a departure of $N_{\rm eff}$ from the standard value of $4/7$ \cite{Lesgourgues:2012uu}.  The presence of such extra degrees of freedom can therefore be tightly constrained using Planck and other observations \cite{Ade:2015xua}. 
Furthermore, structure formation can also be used to constrain the imprints of such light particles. It has been shown recently that the difference between the predictions of the non--local model $R\frac{1}{\square^2}R$ and $\Lambda$CDM is small with respect to the present observational errors \cite{Dirian:2014ara}. 
This is somewhat expected as neutrinos, for instance, being very light with $\sum m_{\nu} \lesssim 0.23 \,{\rm eV}$ induce only sub--leading corrections at a few percent level to $\Lambda$CDM at small scales and a similar conclusion can be anticipated for the even lighter particle in our scenario. 
A more detailed analysis to judge the consistency of this picture will require studying the cosmological evolution of the full model (\ref{ea_R2_Final}) with the exact structure function $\mathcal{F}$ before the simplifying  expansion (\ref{F})  \cite{Maggiore:2016fbn}.
A different microscopic interpretation of the mass scale $m$ has been proposed in \cite{Maggiore:2015rma}.

\vskip 4pt
\paragraph*{\bf Discussion and outlook.}\label{discuss}

In this Letter we have  studied a simple yet elegant scenario which generalizes GR by including leading local $R^2$ and leading non--local $R\frac{1}{\square^2}R$ terms to the action. We have derived the modified Friedman EOM by evaluating the generalized Einstein's equations on an FRW background which are subsequently solved by a combination of analytical and numerical methods. Our main result is that the solution, depicted in Fig. \ref{universe}, exhibits a unified and consistent evolution of the universe that starts from an inflationary regime at very early times and ends in a dark energy phase at the present epoch.

In particular, we have found that the transition between the end of inflation to the radiation epoch is characterized by strong coherent oscillations of the Hubble parameter which may be interpreted as a concrete realization of the reheating phase. These oscillations originate from the competition between $R$ and $R^2$ contributions and are indeed a precise and distinct feature of our scenario (when the theory is treated consistently in the Jordan frame). 
After this transient oscillatory phase, the universe naturally enters the Einstein regime with radiation domination followed by a matter dominated era. 
Eventually, the non--local term starts to dominate and drives the evolution of the universe, behaving as a dynamical and purely gravitational dark energy with $w\leq -1$.
Thus, the model (\ref{action}) is able to describe the evolution of the universe through all the epochs and is characterized by only two new parameters $\xi$ and $m$.
The oscillatory phase it predicts may have characteristic observable imprints which should be analyzed more carefully. 

We emphasize that the unified scenario we studied is not only phenomenological but can be justified
using an effective approach to low energy quantum gravity. The mass scale $m$ can either be identified with the mass of a very light species or with the effective gravitational mass. This connection is not only satisfactory from a theoretical point of view but it also provides a way to reduce the number of free parameters by linking $m$ to the underlying theory.  In conclusion, this scenario represents a unified cosmological model with one parameter less, and is thus characterized by a higher predictive power that can in principle be falsified.

An immediate next step in this direction will be to study the evolution of perturbations in this model and obtain relevant observables to fully construct a viable cosmological model. Although the evolution of perturbations has been studied in the Starobinsky inflation and in the non-local model respectively, it will be very interesting to study them in this unified model which we leave for future work.   
\vskip 2pt
\paragraph*{\bf Acknowledgements.}
 
We would like to thank the anonymous referee for his/her suggestions. 
The CP$^3$-Origins centre is partially funded by the Danish National Research Foundation, grant number DNRF90.



\bibliographystyle{apsrev4-1}
\bibliography{EFT_GR_short_Bibliography}

\begin{thebibliography}{41}%
\makeatletter
\providecommand \@ifxundefined [1]{%
 \@ifx{#1\undefined}
}%
\providecommand \@ifnum [1]{%
 \ifnum #1\expandafter \@firstoftwo
 \else \expandafter \@secondoftwo
 \fi
}%
\providecommand \@ifx [1]{%
 \ifx #1\expandafter \@firstoftwo
 \else \expandafter \@secondoftwo
 \fi
}%
\providecommand \natexlab [1]{#1}%
\providecommand \enquote  [1]{``#1''}%
\providecommand \bibnamefont  [1]{#1}%
\providecommand \bibfnamefont [1]{#1}%
\providecommand \citenamefont [1]{#1}%
\providecommand \href@noop [0]{\@secondoftwo}%
\providecommand \href [0]{\begingroup \@sanitize@url \@href}%
\providecommand \@href[1]{\@@startlink{#1}\@@href}%
\providecommand \@@href[1]{\endgroup#1\@@endlink}%
\providecommand \@sanitize@url [0]{\catcode `\\12\catcode `\$12\catcode
  `\&12\catcode `\#12\catcode `\^12\catcode `\_12\catcode `\%12\relax}%
\providecommand \@@startlink[1]{}%
\providecommand \@@endlink[0]{}%
\providecommand \url  [0]{\begingroup\@sanitize@url \@url }%
\providecommand \@url [1]{\endgroup\@href {#1}{\urlprefix }}%
\providecommand \urlprefix  [0]{URL }%
\providecommand \Eprint [0]{\href }%
\providecommand \doibase [0]{http://dx.doi.org/}%
\providecommand \selectlanguage [0]{\@gobble}%
\providecommand \bibinfo  [0]{\@secondoftwo}%
\providecommand \bibfield  [0]{\@secondoftwo}%
\providecommand \translation [1]{[#1]}%
\providecommand \BibitemOpen [0]{}%
\providecommand \bibitemStop [0]{}%
\providecommand \bibitemNoStop [0]{.\EOS\space}%
\providecommand \EOS [0]{\spacefactor3000\relax}%
\providecommand \BibitemShut  [1]{\csname bibitem#1\endcsname}%
\let\auto@bib@innerbib\@empty
\bibitem [{\citenamefont {Guth}(1981)}]{Guth:1980zm}%
  \BibitemOpen
  \bibfield  {author} {\bibinfo {author} {\bibfnamefont {A.~H.}\ \bibnamefont
  {Guth}},\ }\href {\doibase 10.1103/PhysRevD.23.347} {\bibfield  {journal}
  {\bibinfo  {journal} {Phys. Rev.}\ }\textbf {\bibinfo {volume} {D23}},\
  \bibinfo {pages} {347} (\bibinfo {year} {1981})}\BibitemShut {NoStop}%
\bibitem [{\citenamefont {Ade}\ \emph {et~al.}(2016{\natexlab{a}})\citenamefont
  {Ade} \emph {et~al.}}]{Ade:2015lrj}%
  \BibitemOpen
  \bibfield  {author} {\bibinfo {author} {\bibfnamefont {P.~A.~R.}\
  \bibnamefont {Ade}} \emph {et~al.} (\bibinfo {collaboration} {Planck}),\
  }\href {\doibase 10.1051/0004-6361/201525898} {\bibfield  {journal} {\bibinfo
   {journal} {Astron. Astrophys.}\ }\textbf {\bibinfo {volume} {594}},\
  \bibinfo {pages} {A20} (\bibinfo {year} {2016}{\natexlab{a}})},\ \Eprint
  {http://arxiv.org/abs/1502.02114} {arXiv:1502.02114 [astro-ph.CO]}
  \BibitemShut {NoStop}%
\bibitem [{\citenamefont {Riess}\ \emph {et~al.}(1998)\citenamefont {Riess}
  \emph {et~al.}}]{Riess:1998cb}%
  \BibitemOpen
  \bibfield  {author} {\bibinfo {author} {\bibfnamefont {A.~G.}\ \bibnamefont
  {Riess}} \emph {et~al.} (\bibinfo {collaboration} {Supernova Search Team}),\
  }\href {\doibase 10.1086/300499} {\bibfield  {journal} {\bibinfo  {journal}
  {Astron. J.}\ }\textbf {\bibinfo {volume} {116}},\ \bibinfo {pages} {1009}
  (\bibinfo {year} {1998})},\ \Eprint {http://arxiv.org/abs/astro-ph/9805201}
  {arXiv:astro-ph/9805201 [astro-ph]} \BibitemShut {NoStop}%
\bibitem [{\citenamefont {Perlmutter}\ \emph {et~al.}(1999)\citenamefont
  {Perlmutter} \emph {et~al.}}]{Perlmutter:1998np}%
  \BibitemOpen
  \bibfield  {author} {\bibinfo {author} {\bibfnamefont {S.}~\bibnamefont
  {Perlmutter}} \emph {et~al.} (\bibinfo {collaboration} {Supernova Cosmology
  Project}),\ }\href {\doibase 10.1086/307221} {\bibfield  {journal} {\bibinfo
  {journal} {Astrophys. J.}\ }\textbf {\bibinfo {volume} {517}},\ \bibinfo
  {pages} {565} (\bibinfo {year} {1999})},\ \Eprint
  {http://arxiv.org/abs/astro-ph/9812133} {arXiv:astro-ph/9812133 [astro-ph]}
  \BibitemShut {NoStop}%
\bibitem [{\citenamefont {Peebles}\ and\ \citenamefont
  {Ratra}(2003)}]{Peebles:2002gy}%
  \BibitemOpen
  \bibfield  {author} {\bibinfo {author} {\bibfnamefont {P.~J.~E.}\
  \bibnamefont {Peebles}}\ and\ \bibinfo {author} {\bibfnamefont
  {B.}~\bibnamefont {Ratra}},\ }\href {\doibase 10.1103/RevModPhys.75.559}
  {\bibfield  {journal} {\bibinfo  {journal} {Rev. Mod. Phys.}\ }\textbf
  {\bibinfo {volume} {75}},\ \bibinfo {pages} {559} (\bibinfo {year} {2003})},\
  \Eprint {http://arxiv.org/abs/astro-ph/0207347} {arXiv:astro-ph/0207347
  [astro-ph]} \BibitemShut {NoStop}%
\bibitem [{\citenamefont {Codello}\ and\ \citenamefont
  {Jain}(2016{\natexlab{a}})}]{Codello:2015mba}%
  \BibitemOpen
  \bibfield  {author} {\bibinfo {author} {\bibfnamefont {A.}~\bibnamefont
  {Codello}}\ and\ \bibinfo {author} {\bibfnamefont {R.~K.}\ \bibnamefont
  {Jain}},\ }\href {\doibase 10.1088/0264-9381/33/22/225006} {\bibfield
  {journal} {\bibinfo  {journal} {Class. Quant. Grav.}\ }\textbf {\bibinfo
  {volume} {33}},\ \bibinfo {pages} {225006} (\bibinfo {year}
  {2016}{\natexlab{a}})},\ \Eprint {http://arxiv.org/abs/1507.06308}
  {arXiv:1507.06308 [gr-qc]} \BibitemShut {NoStop}%
\bibitem [{\citenamefont {Codello}\ and\ \citenamefont
  {Jain}(2017)}]{Codello:2015pga}%
  \BibitemOpen
  \bibfield  {author} {\bibinfo {author} {\bibfnamefont {A.}~\bibnamefont
  {Codello}}\ and\ \bibinfo {author} {\bibfnamefont {R.~K.}\ \bibnamefont
  {Jain}},\ }\href {\doibase 10.1088/1361-6382/aa549d} {\bibfield  {journal}
  {\bibinfo  {journal} {Class. Quant. Grav.}\ }\textbf {\bibinfo {volume}
  {34}},\ \bibinfo {pages} {035015} (\bibinfo {year} {2017})},\ \Eprint
  {http://arxiv.org/abs/1507.07829} {arXiv:1507.07829 [astro-ph.CO]}
  \BibitemShut {NoStop}%
\bibitem [{\citenamefont {Codello}\ and\ \citenamefont
  {Jain}(2016{\natexlab{b}})}]{Codello:2016xhm}%
  \BibitemOpen
  \bibfield  {author} {\bibinfo {author} {\bibfnamefont {A.}~\bibnamefont
  {Codello}}\ and\ \bibinfo {author} {\bibfnamefont {R.~K.}\ \bibnamefont
  {Jain}},\ }\href {\doibase 10.1142/S0218271816440235} {\bibfield  {journal}
  {\bibinfo  {journal} {Int. J. Mod. Phys.}\ }\textbf {\bibinfo {volume}
  {D25}},\ \bibinfo {pages} {1644023} (\bibinfo {year} {2016}{\natexlab{b}})},\
  \Eprint {http://arxiv.org/abs/1605.07630} {arXiv:1605.07630 [gr-qc]}
  \BibitemShut {NoStop}%
\bibitem [{\citenamefont {Donoghue}(1994)}]{Donoghue:1994dn}%
  \BibitemOpen
  \bibfield  {author} {\bibinfo {author} {\bibfnamefont {J.~F.}\ \bibnamefont
  {Donoghue}},\ }\href {\doibase 10.1103/PhysRevD.50.3874} {\bibfield
  {journal} {\bibinfo  {journal} {Phys.Rev.}\ }\textbf {\bibinfo {volume}
  {D50}},\ \bibinfo {pages} {3874} (\bibinfo {year} {1994})},\ \Eprint
  {http://arxiv.org/abs/gr-qc/9405057} {arXiv:gr-qc/9405057 [gr-qc]}
  \BibitemShut {NoStop}%
\bibitem [{\citenamefont {Burgess}(2004)}]{Burgess:2003jk}%
  \BibitemOpen
  \bibfield  {author} {\bibinfo {author} {\bibfnamefont {C.}~\bibnamefont
  {Burgess}},\ }\href {\doibase 10.12942/lrr-2004-5} {\bibfield  {journal}
  {\bibinfo  {journal} {Living Rev.Rel.}\ }\textbf {\bibinfo {volume} {7}},\
  \bibinfo {pages} {5} (\bibinfo {year} {2004})},\ \Eprint
  {http://arxiv.org/abs/gr-qc/0311082} {arXiv:gr-qc/0311082 [gr-qc]}
  \BibitemShut {NoStop}%
\bibitem [{\citenamefont {Donoghue}(2012)}]{Donoghue:2012zc}%
  \BibitemOpen
  \bibfield  {author} {\bibinfo {author} {\bibfnamefont {J.~F.}\ \bibnamefont
  {Donoghue}},\ }\bibfield  {booktitle} {\emph {\bibinfo {booktitle}
  {{Proceedings, 6th International School on Field Theory and Gravitation
  (ISFTG 2012)}}},\ }\href {\doibase 10.1063/1.4756964} {\bibfield  {journal}
  {\bibinfo  {journal} {AIP Conf. Proc.}\ }\textbf {\bibinfo {volume} {1483}},\
  \bibinfo {pages} {73} (\bibinfo {year} {2012})},\ \Eprint
  {http://arxiv.org/abs/1209.3511} {arXiv:1209.3511 [gr-qc]} \BibitemShut
  {NoStop}%
\bibitem [{\citenamefont {Donoghue}\ and\ \citenamefont
  {Holstein}(2015)}]{Donoghue:2015hwa}%
  \BibitemOpen
  \bibfield  {author} {\bibinfo {author} {\bibfnamefont {J.~F.}\ \bibnamefont
  {Donoghue}}\ and\ \bibinfo {author} {\bibfnamefont {B.~R.}\ \bibnamefont
  {Holstein}},\ }\href {\doibase 10.1088/0954-3899/42/10/103102} {\bibfield
  {journal} {\bibinfo  {journal} {J. Phys.}\ }\textbf {\bibinfo {volume}
  {G42}},\ \bibinfo {pages} {103102} (\bibinfo {year} {2015})},\ \Eprint
  {http://arxiv.org/abs/1506.00946} {arXiv:1506.00946 [gr-qc]} \BibitemShut
  {NoStop}%
\bibitem [{\citenamefont {Starobinsky}(1980)}]{Starobinsky:1980te}%
  \BibitemOpen
  \bibfield  {author} {\bibinfo {author} {\bibfnamefont {A.~A.}\ \bibnamefont
  {Starobinsky}},\ }\href {\doibase 10.1016/0370-2693(80)90670-X} {\bibfield
  {journal} {\bibinfo  {journal} {Phys. Lett.}\ }\textbf {\bibinfo {volume}
  {B91}},\ \bibinfo {pages} {99} (\bibinfo {year} {1980})}\BibitemShut
  {NoStop}%
\bibitem [{\citenamefont {Mukhanov}\ and\ \citenamefont
  {Chibisov}(1981)}]{Mukhanov:1981xt}%
  \BibitemOpen
  \bibfield  {author} {\bibinfo {author} {\bibfnamefont {V.~F.}\ \bibnamefont
  {Mukhanov}}\ and\ \bibinfo {author} {\bibfnamefont {G.~V.}\ \bibnamefont
  {Chibisov}},\ }\href@noop {} {\bibfield  {journal} {\bibinfo  {journal} {JETP
  Lett.}\ }\textbf {\bibinfo {volume} {33}},\ \bibinfo {pages} {532} (\bibinfo
  {year} {1981})},\ \bibinfo {note} {[Pisma Zh. Eksp. Teor.
  Fiz.33,549(1981)]}\BibitemShut {NoStop}%
\bibitem [{\citenamefont {Starobinsky}(1983)}]{Starobinsky:1983zz}%
  \BibitemOpen
  \bibfield  {author} {\bibinfo {author} {\bibfnamefont {A.~A.}\ \bibnamefont
  {Starobinsky}},\ }\href@noop {} {\bibfield  {journal} {\bibinfo  {journal}
  {Sov. Astron. Lett.}\ }\textbf {\bibinfo {volume} {9}},\ \bibinfo {pages}
  {302} (\bibinfo {year} {1983})}\BibitemShut {NoStop}%
\bibitem [{\citenamefont {Kehagias}\ \emph {et~al.}(2014)\citenamefont
  {Kehagias}, \citenamefont {Moradinezhad~Dizgah},\ and\ \citenamefont
  {Riotto}}]{Kehagias:2013mya}%
  \BibitemOpen
  \bibfield  {author} {\bibinfo {author} {\bibfnamefont {A.}~\bibnamefont
  {Kehagias}}, \bibinfo {author} {\bibfnamefont {A.}~\bibnamefont
  {Moradinezhad~Dizgah}}, \ and\ \bibinfo {author} {\bibfnamefont
  {A.}~\bibnamefont {Riotto}},\ }\href {\doibase 10.1103/PhysRevD.89.043527}
  {\bibfield  {journal} {\bibinfo  {journal} {Phys. Rev.}\ }\textbf {\bibinfo
  {volume} {D89}},\ \bibinfo {pages} {043527} (\bibinfo {year} {2014})},\
  \Eprint {http://arxiv.org/abs/1312.1155} {arXiv:1312.1155 [hep-th]}
  \BibitemShut {NoStop}%
\bibitem [{\citenamefont {Martin}\ \emph {et~al.}(2014)\citenamefont {Martin},
  \citenamefont {Ringeval},\ and\ \citenamefont {Vennin}}]{Martin:2013tda}%
  \BibitemOpen
  \bibfield  {author} {\bibinfo {author} {\bibfnamefont {J.}~\bibnamefont
  {Martin}}, \bibinfo {author} {\bibfnamefont {C.}~\bibnamefont {Ringeval}}, \
  and\ \bibinfo {author} {\bibfnamefont {V.}~\bibnamefont {Vennin}},\ }\href
  {\doibase 10.1016/j.dark.2014.01.003} {\bibfield  {journal} {\bibinfo
  {journal} {Phys. Dark Univ.}\ }\textbf {\bibinfo {volume} {5-6}},\ \bibinfo
  {pages} {75} (\bibinfo {year} {2014})},\ \Eprint
  {http://arxiv.org/abs/1303.3787} {arXiv:1303.3787 [astro-ph.CO]} \BibitemShut
  {NoStop}%
\bibitem [{\citenamefont {Codello}\ \emph {et~al.}(2015)\citenamefont
  {Codello}, \citenamefont {Joergensen}, \citenamefont {Sannino},\ and\
  \citenamefont {Svendsen}}]{Codello:2014sua}%
  \BibitemOpen
  \bibfield  {author} {\bibinfo {author} {\bibfnamefont {A.}~\bibnamefont
  {Codello}}, \bibinfo {author} {\bibfnamefont {J.}~\bibnamefont {Joergensen}},
  \bibinfo {author} {\bibfnamefont {F.}~\bibnamefont {Sannino}}, \ and\
  \bibinfo {author} {\bibfnamefont {O.}~\bibnamefont {Svendsen}},\ }\href
  {\doibase 10.1007/JHEP02(2015)050} {\bibfield  {journal} {\bibinfo  {journal}
  {JHEP}\ }\textbf {\bibinfo {volume} {1502}},\ \bibinfo {pages} {050}
  (\bibinfo {year} {2015})},\ \Eprint {http://arxiv.org/abs/1404.3558}
  {arXiv:1404.3558 [hep-ph]} \BibitemShut {NoStop}%
\bibitem [{\citenamefont {Maggiore}\ and\ \citenamefont
  {Mancarella}(2014)}]{Maggiore:2014sia}%
  \BibitemOpen
  \bibfield  {author} {\bibinfo {author} {\bibfnamefont {M.}~\bibnamefont
  {Maggiore}}\ and\ \bibinfo {author} {\bibfnamefont {M.}~\bibnamefont
  {Mancarella}},\ }\href {\doibase 10.1103/PhysRevD.90.023005} {\bibfield
  {journal} {\bibinfo  {journal} {Phys. Rev.}\ }\textbf {\bibinfo {volume}
  {D90}},\ \bibinfo {pages} {023005} (\bibinfo {year} {2014})},\ \Eprint
  {http://arxiv.org/abs/1402.0448} {arXiv:1402.0448 [hep-th]} \BibitemShut
  {NoStop}%
\bibitem [{\citenamefont {Dirian}\ \emph {et~al.}(2014)\citenamefont {Dirian},
  \citenamefont {Foffa}, \citenamefont {Khosravi}, \citenamefont {Kunz},\ and\
  \citenamefont {Maggiore}}]{Dirian:2014ara}%
  \BibitemOpen
  \bibfield  {author} {\bibinfo {author} {\bibfnamefont {Y.}~\bibnamefont
  {Dirian}}, \bibinfo {author} {\bibfnamefont {S.}~\bibnamefont {Foffa}},
  \bibinfo {author} {\bibfnamefont {N.}~\bibnamefont {Khosravi}}, \bibinfo
  {author} {\bibfnamefont {M.}~\bibnamefont {Kunz}}, \ and\ \bibinfo {author}
  {\bibfnamefont {M.}~\bibnamefont {Maggiore}},\ }\href {\doibase
  10.1088/1475-7516/2014/06/033} {\bibfield  {journal} {\bibinfo  {journal}
  {JCAP}\ }\textbf {\bibinfo {volume} {1406}},\ \bibinfo {pages} {033}
  (\bibinfo {year} {2014})},\ \Eprint {http://arxiv.org/abs/1403.6068}
  {arXiv:1403.6068 [astro-ph.CO]} \BibitemShut {NoStop}%
\bibitem [{\citenamefont {Maggiore}(2015)}]{Maggiore:2015rma}%
  \BibitemOpen
  \bibfield  {author} {\bibinfo {author} {\bibfnamefont {M.}~\bibnamefont
  {Maggiore}},\ }\href@noop {} {\  (\bibinfo {year} {2015})},\ \Eprint
  {http://arxiv.org/abs/1506.06217} {arXiv:1506.06217 [hep-th]} \BibitemShut
  {NoStop}%
\bibitem [{\citenamefont {Cusin}\ \emph
  {et~al.}(2016{\natexlab{a}})\citenamefont {Cusin}, \citenamefont {Foffa},
  \citenamefont {Maggiore},\ and\ \citenamefont {Mancarella}}]{Cusin:2015rex}%
  \BibitemOpen
  \bibfield  {author} {\bibinfo {author} {\bibfnamefont {G.}~\bibnamefont
  {Cusin}}, \bibinfo {author} {\bibfnamefont {S.}~\bibnamefont {Foffa}},
  \bibinfo {author} {\bibfnamefont {M.}~\bibnamefont {Maggiore}}, \ and\
  \bibinfo {author} {\bibfnamefont {M.}~\bibnamefont {Mancarella}},\ }\href
  {\doibase 10.1103/PhysRevD.93.043006} {\bibfield  {journal} {\bibinfo
  {journal} {Phys. Rev.}\ }\textbf {\bibinfo {volume} {D93}},\ \bibinfo {pages}
  {043006} (\bibinfo {year} {2016}{\natexlab{a}})},\ \Eprint
  {http://arxiv.org/abs/1512.06373} {arXiv:1512.06373 [hep-th]} \BibitemShut
  {NoStop}%
\bibitem [{\citenamefont {Cusin}\ \emph
  {et~al.}(2016{\natexlab{b}})\citenamefont {Cusin}, \citenamefont {Foffa},
  \citenamefont {Maggiore},\ and\ \citenamefont {Mancarella}}]{Cusin:2016nzi}%
  \BibitemOpen
  \bibfield  {author} {\bibinfo {author} {\bibfnamefont {G.}~\bibnamefont
  {Cusin}}, \bibinfo {author} {\bibfnamefont {S.}~\bibnamefont {Foffa}},
  \bibinfo {author} {\bibfnamefont {M.}~\bibnamefont {Maggiore}}, \ and\
  \bibinfo {author} {\bibfnamefont {M.}~\bibnamefont {Mancarella}},\ }\href
  {\doibase 10.1103/PhysRevD.93.083008} {\bibfield  {journal} {\bibinfo
  {journal} {Phys. Rev.}\ }\textbf {\bibinfo {volume} {D93}},\ \bibinfo {pages}
  {083008} (\bibinfo {year} {2016}{\natexlab{b}})},\ \Eprint
  {http://arxiv.org/abs/1602.01078} {arXiv:1602.01078 [hep-th]} \BibitemShut
  {NoStop}%
\bibitem [{\citenamefont {Espriu}\ \emph {et~al.}(2005)\citenamefont {Espriu},
  \citenamefont {Multamaki},\ and\ \citenamefont {Vagenas}}]{Espriu:2005qn}%
  \BibitemOpen
  \bibfield  {author} {\bibinfo {author} {\bibfnamefont {D.}~\bibnamefont
  {Espriu}}, \bibinfo {author} {\bibfnamefont {T.}~\bibnamefont {Multamaki}}, \
  and\ \bibinfo {author} {\bibfnamefont {E.~C.}\ \bibnamefont {Vagenas}},\
  }\href {\doibase 10.1016/j.physletb.2005.09.033} {\bibfield  {journal}
  {\bibinfo  {journal} {Phys. Lett.}\ }\textbf {\bibinfo {volume} {B628}},\
  \bibinfo {pages} {197} (\bibinfo {year} {2005})},\ \Eprint
  {http://arxiv.org/abs/gr-qc/0503033} {arXiv:gr-qc/0503033 [gr-qc]}
  \BibitemShut {NoStop}%
\bibitem [{\citenamefont {Deser}\ and\ \citenamefont
  {Woodard}(2007)}]{Deser:2007jk}%
  \BibitemOpen
  \bibfield  {author} {\bibinfo {author} {\bibfnamefont {S.}~\bibnamefont
  {Deser}}\ and\ \bibinfo {author} {\bibfnamefont {R.~P.}\ \bibnamefont
  {Woodard}},\ }\href {\doibase 10.1103/PhysRevLett.99.111301} {\bibfield
  {journal} {\bibinfo  {journal} {Phys. Rev. Lett.}\ }\textbf {\bibinfo
  {volume} {99}},\ \bibinfo {pages} {111301} (\bibinfo {year} {2007})},\
  \Eprint {http://arxiv.org/abs/0706.2151} {arXiv:0706.2151 [astro-ph]}
  \BibitemShut {NoStop}%
\bibitem [{\citenamefont {Nojiri}\ and\ \citenamefont
  {Odintsov}(2008)}]{Nojiri:2007uq}%
  \BibitemOpen
  \bibfield  {author} {\bibinfo {author} {\bibfnamefont {S.}~\bibnamefont
  {Nojiri}}\ and\ \bibinfo {author} {\bibfnamefont {S.~D.}\ \bibnamefont
  {Odintsov}},\ }\href {\doibase 10.1016/j.physletb.2007.12.001} {\bibfield
  {journal} {\bibinfo  {journal} {Phys. Lett.}\ }\textbf {\bibinfo {volume}
  {B659}},\ \bibinfo {pages} {821} (\bibinfo {year} {2008})},\ \Eprint
  {http://arxiv.org/abs/0708.0924} {arXiv:0708.0924 [hep-th]} \BibitemShut
  {NoStop}%
\bibitem [{\citenamefont {Koivisto}(2008)}]{Koivisto:2008xfa}%
  \BibitemOpen
  \bibfield  {author} {\bibinfo {author} {\bibfnamefont {T.}~\bibnamefont
  {Koivisto}},\ }\href {\doibase 10.1103/PhysRevD.77.123513} {\bibfield
  {journal} {\bibinfo  {journal} {Phys. Rev.}\ }\textbf {\bibinfo {volume}
  {D77}},\ \bibinfo {pages} {123513} (\bibinfo {year} {2008})},\ \Eprint
  {http://arxiv.org/abs/0803.3399} {arXiv:0803.3399 [gr-qc]} \BibitemShut
  {NoStop}%
\bibitem [{\citenamefont {Nojiri}\ \emph {et~al.}(2011)\citenamefont {Nojiri},
  \citenamefont {Odintsov}, \citenamefont {Sasaki},\ and\ \citenamefont
  {Zhang}}]{Nojiri:2010pw}%
  \BibitemOpen
  \bibfield  {author} {\bibinfo {author} {\bibfnamefont {S.}~\bibnamefont
  {Nojiri}}, \bibinfo {author} {\bibfnamefont {S.~D.}\ \bibnamefont
  {Odintsov}}, \bibinfo {author} {\bibfnamefont {M.}~\bibnamefont {Sasaki}}, \
  and\ \bibinfo {author} {\bibfnamefont {Y.-l.}\ \bibnamefont {Zhang}},\ }\href
  {\doibase 10.1016/j.physletb.2010.12.035} {\bibfield  {journal} {\bibinfo
  {journal} {Phys. Lett.}\ }\textbf {\bibinfo {volume} {B696}},\ \bibinfo
  {pages} {278} (\bibinfo {year} {2011})},\ \Eprint
  {http://arxiv.org/abs/1010.5375} {arXiv:1010.5375 [gr-qc]} \BibitemShut
  {NoStop}%
\bibitem [{\citenamefont {Biswas}\ \emph {et~al.}(2010)\citenamefont {Biswas},
  \citenamefont {Koivisto},\ and\ \citenamefont {Mazumdar}}]{Biswas:2010zk}%
  \BibitemOpen
  \bibfield  {author} {\bibinfo {author} {\bibfnamefont {T.}~\bibnamefont
  {Biswas}}, \bibinfo {author} {\bibfnamefont {T.}~\bibnamefont {Koivisto}}, \
  and\ \bibinfo {author} {\bibfnamefont {A.}~\bibnamefont {Mazumdar}},\ }\href
  {\doibase 10.1088/1475-7516/2010/11/008} {\bibfield  {journal} {\bibinfo
  {journal} {JCAP}\ }\textbf {\bibinfo {volume} {1011}},\ \bibinfo {pages}
  {008} (\bibinfo {year} {2010})},\ \Eprint {http://arxiv.org/abs/1005.0590}
  {arXiv:1005.0590 [hep-th]} \BibitemShut {NoStop}%
\bibitem [{\citenamefont {Nojiri}\ and\ \citenamefont
  {Odintsov}(2011)}]{Nojiri:2010wj}%
  \BibitemOpen
  \bibfield  {author} {\bibinfo {author} {\bibfnamefont {S.}~\bibnamefont
  {Nojiri}}\ and\ \bibinfo {author} {\bibfnamefont {S.~D.}\ \bibnamefont
  {Odintsov}},\ }\href {\doibase 10.1016/j.physrep.2011.04.001} {\bibfield
  {journal} {\bibinfo  {journal} {Phys. Rept.}\ }\textbf {\bibinfo {volume}
  {505}},\ \bibinfo {pages} {59} (\bibinfo {year} {2011})},\ \Eprint
  {http://arxiv.org/abs/1011.0544} {arXiv:1011.0544 [gr-qc]} \BibitemShut
  {NoStop}%
\bibitem [{\citenamefont {Zhang}\ and\ \citenamefont
  {Sasaki}(2012)}]{Zhang:2011uv}%
  \BibitemOpen
  \bibfield  {author} {\bibinfo {author} {\bibfnamefont {Y.-l.}\ \bibnamefont
  {Zhang}}\ and\ \bibinfo {author} {\bibfnamefont {M.}~\bibnamefont {Sasaki}},\
  }\href {\doibase 10.1142/S021827181250006X} {\bibfield  {journal} {\bibinfo
  {journal} {Int. J. Mod. Phys.}\ }\textbf {\bibinfo {volume} {D21}},\ \bibinfo
  {pages} {1250006} (\bibinfo {year} {2012})},\ \Eprint
  {http://arxiv.org/abs/1108.2112} {arXiv:1108.2112 [gr-qc]} \BibitemShut
  {NoStop}%
\bibitem [{\citenamefont {Modesto}\ and\ \citenamefont
  {Tsujikawa}(2013)}]{Modesto:2013jea}%
  \BibitemOpen
  \bibfield  {author} {\bibinfo {author} {\bibfnamefont {L.}~\bibnamefont
  {Modesto}}\ and\ \bibinfo {author} {\bibfnamefont {S.}~\bibnamefont
  {Tsujikawa}},\ }\href {\doibase 10.1016/j.physletb.2013.10.037} {\bibfield
  {journal} {\bibinfo  {journal} {Phys.Lett.}\ }\textbf {\bibinfo {volume}
  {B727}},\ \bibinfo {pages} {48} (\bibinfo {year} {2013})},\ \Eprint
  {http://arxiv.org/abs/1307.6968} {arXiv:1307.6968 [hep-th]} \BibitemShut
  {NoStop}%
\bibitem [{\citenamefont {Donoghue}\ and\ \citenamefont
  {El-Menoufi}(2014)}]{Donoghue:2014yha}%
  \BibitemOpen
  \bibfield  {author} {\bibinfo {author} {\bibfnamefont {J.~F.}\ \bibnamefont
  {Donoghue}}\ and\ \bibinfo {author} {\bibfnamefont {B.~K.}\ \bibnamefont
  {El-Menoufi}},\ }\href {\doibase 10.1103/PhysRevD.89.104062} {\bibfield
  {journal} {\bibinfo  {journal} {Phys.Rev.}\ }\textbf {\bibinfo {volume}
  {D89}},\ \bibinfo {pages} {104062} (\bibinfo {year} {2014})},\ \Eprint
  {http://arxiv.org/abs/1402.3252} {arXiv:1402.3252 [gr-qc]} \BibitemShut
  {NoStop}%
\bibitem [{\citenamefont {Modesto}\ and\ \citenamefont
  {Rachwal}(2014)}]{Modesto:2014lga}%
  \BibitemOpen
  \bibfield  {author} {\bibinfo {author} {\bibfnamefont {L.}~\bibnamefont
  {Modesto}}\ and\ \bibinfo {author} {\bibfnamefont {L.}~\bibnamefont
  {Rachwal}},\ }\href {\doibase 10.1016/j.nuclphysb.2014.10.015} {\bibfield
  {journal} {\bibinfo  {journal} {Nucl.Phys.}\ }\textbf {\bibinfo {volume}
  {B889}},\ \bibinfo {pages} {228} (\bibinfo {year} {2014})},\ \Eprint
  {http://arxiv.org/abs/1407.8036} {arXiv:1407.8036 [hep-th]} \BibitemShut
  {NoStop}%
\bibitem [{\citenamefont {Zhang}\ \emph {et~al.}(2016)\citenamefont {Zhang},
  \citenamefont {Koyama}, \citenamefont {Sasaki},\ and\ \citenamefont
  {Zhao}}]{Zhang:2016ykx}%
  \BibitemOpen
  \bibfield  {author} {\bibinfo {author} {\bibfnamefont {Y.-l.}\ \bibnamefont
  {Zhang}}, \bibinfo {author} {\bibfnamefont {K.}~\bibnamefont {Koyama}},
  \bibinfo {author} {\bibfnamefont {M.}~\bibnamefont {Sasaki}}, \ and\ \bibinfo
  {author} {\bibfnamefont {G.-B.}\ \bibnamefont {Zhao}},\ }\href {\doibase
  10.1007/JHEP03(2016)039} {\bibfield  {journal} {\bibinfo  {journal} {JHEP}\
  }\textbf {\bibinfo {volume} {03}},\ \bibinfo {pages} {039} (\bibinfo {year}
  {2016})},\ \Eprint {http://arxiv.org/abs/1601.03808} {arXiv:1601.03808
  [hep-th]} \BibitemShut {NoStop}%
\bibitem [{\citenamefont {Shapiro}(2008)}]{Shapiro:2008sf}%
  \BibitemOpen
  \bibfield  {author} {\bibinfo {author} {\bibfnamefont {I.~L.}\ \bibnamefont
  {Shapiro}},\ }\href {\doibase 10.1088/0264-9381/25/10/103001} {\bibfield
  {journal} {\bibinfo  {journal} {Class. Quant. Grav.}\ }\textbf {\bibinfo
  {volume} {25}},\ \bibinfo {pages} {103001} (\bibinfo {year} {2008})},\
  \Eprint {http://arxiv.org/abs/0801.0216} {arXiv:0801.0216 [gr-qc]}
  \BibitemShut {NoStop}%
\bibitem [{\citenamefont {Mottola}(2010)}]{Mottola:2010gp}%
  \BibitemOpen
  \bibfield  {author} {\bibinfo {author} {\bibfnamefont {E.}~\bibnamefont
  {Mottola}},\ }\href@noop {} {\bibfield  {journal} {\bibinfo  {journal} {Acta
  Phys.Polon.}\ }\textbf {\bibinfo {volume} {B41}},\ \bibinfo {pages} {2031}
  (\bibinfo {year} {2010})},\ \Eprint {http://arxiv.org/abs/1008.5006}
  {arXiv:1008.5006 [gr-qc]} \BibitemShut {NoStop}%
\bibitem [{\citenamefont {Netto}\ \emph {et~al.}(2016)\citenamefont {Netto},
  \citenamefont {Pelinson}, \citenamefont {Shapiro},\ and\ \citenamefont
  {Starobinsky}}]{Netto:2015cba}%
  \BibitemOpen
  \bibfield  {author} {\bibinfo {author} {\bibfnamefont {T.~d.~P.}\
  \bibnamefont {Netto}}, \bibinfo {author} {\bibfnamefont {A.~M.}\ \bibnamefont
  {Pelinson}}, \bibinfo {author} {\bibfnamefont {I.~L.}\ \bibnamefont
  {Shapiro}}, \ and\ \bibinfo {author} {\bibfnamefont {A.~A.}\ \bibnamefont
  {Starobinsky}},\ }\href {\doibase 10.1140/epjc/s10052-016-4390-4} {\bibfield
  {journal} {\bibinfo  {journal} {Eur. Phys. J.}\ }\textbf {\bibinfo {volume}
  {C76}},\ \bibinfo {pages} {544} (\bibinfo {year} {2016})},\ \Eprint
  {http://arxiv.org/abs/1509.08882} {arXiv:1509.08882 [hep-th]} \BibitemShut
  {NoStop}%
\bibitem [{\citenamefont {Lesgourgues}\ and\ \citenamefont
  {Pastor}(2012)}]{Lesgourgues:2012uu}%
  \BibitemOpen
  \bibfield  {author} {\bibinfo {author} {\bibfnamefont {J.}~\bibnamefont
  {Lesgourgues}}\ and\ \bibinfo {author} {\bibfnamefont {S.}~\bibnamefont
  {Pastor}},\ }\href {\doibase 10.1155/2012/608515} {\bibfield  {journal}
  {\bibinfo  {journal} {Adv. High Energy Phys.}\ }\textbf {\bibinfo {volume}
  {2012}},\ \bibinfo {pages} {608515} (\bibinfo {year} {2012})},\ \Eprint
  {http://arxiv.org/abs/1212.6154} {arXiv:1212.6154 [hep-ph]} \BibitemShut
  {NoStop}%
\bibitem [{\citenamefont {Ade}\ \emph {et~al.}(2016{\natexlab{b}})\citenamefont
  {Ade} \emph {et~al.}}]{Ade:2015xua}%
  \BibitemOpen
  \bibfield  {author} {\bibinfo {author} {\bibfnamefont {P.~A.~R.}\
  \bibnamefont {Ade}} \emph {et~al.} (\bibinfo {collaboration} {Planck}),\
  }\href {\doibase 10.1051/0004-6361/201525830} {\bibfield  {journal} {\bibinfo
   {journal} {Astron. Astrophys.}\ }\textbf {\bibinfo {volume} {594}},\
  \bibinfo {pages} {A13} (\bibinfo {year} {2016}{\natexlab{b}})},\ \Eprint
  {http://arxiv.org/abs/1502.01589} {arXiv:1502.01589 [astro-ph.CO]}
  \BibitemShut {NoStop}%
\bibitem [{\citenamefont {Maggiore}(2016)}]{Maggiore:2016fbn}%
  \BibitemOpen
  \bibfield  {author} {\bibinfo {author} {\bibfnamefont {M.}~\bibnamefont
  {Maggiore}},\ }\href {\doibase 10.1103/PhysRevD.93.063008} {\bibfield
  {journal} {\bibinfo  {journal} {Phys. Rev.}\ }\textbf {\bibinfo {volume}
  {D93}},\ \bibinfo {pages} {063008} (\bibinfo {year} {2016})},\ \Eprint
  {http://arxiv.org/abs/1603.01515} {arXiv:1603.01515 [hep-th]} \BibitemShut
  {NoStop}%
\end{thebibliography}%


\end{document}